\documentclass[12pt]{revtex4}
\usepackage{epsfig}
\usepackage{amsfonts}
\usepackage{amsmath}
\usepackage{amssymb}
\usepackage{graphicx}
\usepackage{mathptmx}

\def\be{\begin{eqnarray}}
\def\en{\end{eqnarray}}
\def\non{\nonumber}
\def\vp{\varepsilon}
\def\ra{\rangle}
\def\la{\langle}

\def\ov{\overline}
\begin{document}
\title{Branching Fractions and $CP$ Asymmetries of $B \to K_0^*(1430)\rho$ and
$B \to K_0^*(1430)\phi $ Decays  in the Family Nonuniversal
$Z^\prime$ Model}
\author{Ying Li}
\email{liying@ytu.edu.cn}
\author{En-Lei Wang }
\date{\today}
\affiliation{Department of Physics, Yantai University, Yantai
264-005, China}
\begin{abstract}
In this work, within the QCD factorization approach, we investigate
the branching fractions and $CP$ asymmetries of decays $B \to
K_0^*(1430)\rho$ and $B \to K_0^*(1430)\phi$ under two different
scenarios both in the standard model and  the family nonuniversal
$Z^\prime$ model. We find that the annihilation terms paly crucial
roles in these decays and lead to the main uncertainties. For decays
$B^- \to \overline  K_0^{*-} (1430)\rho^0 (\omega)$, the new $Z^\prime$ boson
could change branching fractions remarkably. However, for other
decays, its contribution might be clouded by large uncertainties from annihilations. Unfortunately, neither the standard model nor $Z^\prime$ model  can reproduce all experimental data under one certain scenario. We
also noted that the $CP$ asymmetries of  $B^-\to \overline K_0^{*-}(1430)
\rho^0 (\omega)$ could be used to identify the
$K_0^*(1430)$ meson and search for the new physics contribution.
\end{abstract}
\pacs{} \maketitle
\section{Introduction}\label{sec:1}
The study of $B$ meson rare decays is a crucial tool in testing the
fundamental interactions among elementary particles, exploring the
origin of $CP$ violation, and searching for possible new physics
(NP) beyond the standard model (SM). Theoretically and
experimentally, such kind of research has been conducted in great
detail, especially in the weak interactions of $B$ meson. In
particular, the processes induced by flavor-changing neutral-current
(FCNC) only occur at the loop level in SM, and are therefore a very sensitive probe of NP beyond SM . Already, FCNC processes have been explored mainly  in the $B_q -\bar B_q$ mixing  and the
semi-leptonic weak decays, which permit a clean theoretical
description. So far, the charmless hadronic $B$ meson decays induced
by FCNC have also been studied extensively, such as $B \to K\pi,
K^{(*)}\phi$ and $K^{(*)}\eta^{(\prime)}$ decays. In the past few
years, the new physics effect in these decays have also been studied
widely, such as in supersymmetry model, two-Higgs doublet model,
$Z^\prime$ model, the forth generation model, extra dimension
models, and so on (see review in \cite{Cheng:2009xz} and references
therein).

In order to search for effect of NP in the nonleptonic $B$ decays, most
theoretical studies are focused on $B \to PP$, $PV$ or $VV$ in the
past few years. But, the studies of decay modes involving a scalar
meson are relatively few, because the underlying structure of the
scalar mesons is not well established in theoretical side. To
describe the component of the scalar mesons, there are usually two
possible scenarios (S1 and S2) according to the QCD sum rule method
\cite{Minkowski:1998mf}: (i) In S1, we treat scalars above 1 GeV as
the first excited states, while the scalars under 1 GeV are regarded
as the low lying states; (ii) In S2, the scalars above 1 GeV are
viewed as the ground states, and light scalars are four-quark bound
states or hybrid states. Under these two scenarios, many special
decays have been examined within  the QCD factorization
(QCDF) approach \cite{Cheng:2005nb,Cheng:2007st}  or the  perturbative
QCD approach (pQCD)
\cite{Chen:2005cx,Chen:2007qj,shen,Kim:2009dg,Liu:2009xm,Liu:2010zg,
Zhang:2010kw,Zhang:2010af,Liu:2011pe,Li:2011kw}. However, because of large uncertainties in SM, 
the NP effects in these decays are rarely studied.

Very recently, BaBar collaboration reported their first branching
fraction measurements for the decays $B \to K_0^*(1430)\rho$ that
are induced by FCNC \cite{Lees:2011dq}:
\begin{eqnarray}
\label{eq:data1}
 {Br}( B^0\to K_0^{*0}(1430)\rho^0 ) &=& (27\pm 4\pm 2\pm 3)\times
 10^{-6};\\
 {Br}( B^0\to  K_0^{*+}(1430)\rho^-) &=& (28\pm 10\pm 5\pm 3)\times
 10^{-6}.
\end{eqnarray}
The above results are inconsistent with the pQCD predictions
\cite{Zhang:2010af} in most cases. Moreover, these results are
somewhat much lower than the QCDF predictions \cite{Cheng:2007st}
but are consistent with QCDF within rather large uncertainties. For
$B \to K_0^*(1430)\phi$, BaBar collaborator also updated their
results \cite{Aubert:2006uk, 2008zzd} in ref.~\cite{Gao:2009zz}:
\begin{eqnarray}
\label{eq:data2}
 {Br}(B^0\to  K_0^{*0}(1430)\phi) &=& (4.3\pm0.6\pm0.4)\times
 10^{-6};\\
 {Br}( B^\pm\to  K_0^{*\pm}(1430)\phi) &=& (7.0\pm1.3\pm0.9)\times
 10^{-6}.
\end{eqnarray}
Both QCDF and pQCD calculation of above modes have also been
presented in Refs.\cite{Cheng:2007st,Kim:2009dg}, and the predicted
central values of $B^0\to \phi  K_0^{*0}(1430)$ deviate from the
experimental data, though they can be also accommodated within very
large theoretical errors. In the following, $K_0ˆ*(1430) $ is denoted as $K_0^*$
in some places for convenience. 

The predictions of SM cannot agree the data convincingly, which
gives us possible hints on physics beyond SM. It is our purpose of
this work to show that a new physics effect of similar size can be
obtained from some models with an extra spin-1 $Z^\prime$ bosons,
which are known to naturally exist in some well-motivated extensions
of the SM \cite{Langacker:2008yv}. Interesting phenomena arise when
the $Z^\prime$ couplings to physical fermion eigenstates are
nondiagonal, which could be realized in the $E_6$ models
\cite{Nardi:1992nq}, string models \cite{Chaudhuri:1994cd} and some
grand unified theories \cite{GUTs}. For example, in the super string
model advocated by Chaudhuri {\it et.al.}\cite{Chaudhuri:1994cd}, it
is possible to have family nonuniversal $Z^\prime$ couplings,
because of different constructions of the different families. It
also should be note that in such a model, called the family
nonuniversal $Z^\prime$ model, the nonuniversal couplings could lead
to FCNCs at the tree level as well as introduce new weak phases
\cite{Langacker:2000ju}, which could explain the $CP$ asymmetries in
the current high energy experiments. In fact, the effects of
$Z^\prime$ models  have been studied extensively in the low energy
flavor physics phenomena, such as neutral mesons mixing, $B$ meson
decays, single top production and lepton decays
\cite{Langacker:2000ju,Barger:2003hg,Kim:2012rp,
Barger:2009hn,Cheung:2006tm,Chang:2009wt,Hua:2010zy,Arhrib:2006sg,Chiang:2011cv}.

In this current work, we shall adopt the QCD factorization approach
\cite{Beneke:1999br} to evaluate the relevant hadronic matrix
elements of $B$ decays, since it is a systematic framework to calculate these matrix
elements from QCD theory, and holds in the heavy quark limit $m_b
\to \infty$ and the heavy quark symmetry.  In such calculations, one
requires the additional knowledge about form factors of $B$ meson to
the scalar or the vector transitions. This problem, being a part of
the nonperturbative sector of QCD, lacks a precise solution. To the
best of our knowledge, a number of different approaches had been
used to calculate the form factors of $B \to S$ decays, such as QCD
sum rule \cite{Yang:2005bv,Aliev:2007rq}, light-cone QCD sum rule
\cite{Wang:2008da,Sun:2010nv}, perturbative QCD approach
\cite{Li:2008tk} and covariant light front quark model (cLFQM)
\cite{Cheng:2003sm}. Among them, the form factors of the cLFQM are
first calculated in the spacelike region and their momentum
dependence is fitted to a 3-parameter form. This parameterization is
then analytically continued to the timelike region to determine the
physical form factors at $q^2\geq 0$. Moreover, for these form
factors both the heavy quark limit and heavy quark symmetry are
satisfied. For that reason, we will use the results of cLFQM
\cite{Cheng:2003sm} in the following calculations.

For comparison, $B \to K_0^*\rho$ and $ K_0^*\phi$ decays in SM  should be
reinvestigated in Section.\ref{sec:2}.  In Section.\ref{sec:3}, we
will review the family nonuniversal $Z^\prime$ model briefly and show the effect of
$Z^\prime$ to decay modes we are considering. In Section.\ref{sec:4}, we
will present our numerical results and discussions in great detail. At last, we will summrize this work in
Section.\ref{sec:5} .
\section{Revisiting $B \to \rho K_0^*(1430)$ and $B \to \phi K_0^*(1430)$
decays within the QCDF framework}\label{sec:2}
To proceed, we  discuss the decay constants of the scalar
meson. Unlike pseudoscalar meson, each scalar meson has two decay constants, the
vector decay constant $f_S$ and the scale-dependent scalar decay
constant $\bar f_S$ namely,  which are defined as:
\begin{eqnarray}
\langle S(p)|\bar q_2\gamma_\mu q_1|0\ra=f_Sp_\mu,
\,\,\,\,\,\,\,\,\,\,\, \langle S(p)|\bar q_2 q_1|0\ra=m_S\bar {f_S},
\end{eqnarray}
and they are related by the equation of motion:
\begin{eqnarray} \label{eq:EOM}
 f_S=\frac{m_2(\mu)-m_1(\mu)}{m_S} \bar f_S,
\end{eqnarray}
where $m_{2}$ and $m_{1}$ are the running current quark masses.
Therefore, the vector decay constant is much  smaller than the
scalar one. As for the vector meson, the two kinds of decay
constants are also given by \cite{Ball:2004rg}
\begin{eqnarray}
\langle V(p)|\bar q_2\gamma_\mu q_1|0\ra=f_Vm_V\vp_\mu^*,
\,\,\,\,\,\,\,\,\,\,\, \la V(p,\vp^*)|\bar
q\sigma_{\mu\nu}q'|0\ra=f_V^\bot(p_\mu
 \vp_\nu^*-p_\nu\vp_\mu^*).
\end{eqnarray}

The twist-2 and twist-3 light-cone distribution amplitudes (LCDAs)
of scalar mesons, $\phi_{S}(x)$, $\phi^s_{S}(x)$ and
$\phi^{\sigma}_{S}(x)$ respect the normalization conditions:
\begin{eqnarray}
\int^1_0dx\phi_{S}(x)=\frac{f_{S}}{2\sqrt{6}},\,\,\,\,\,\,
\int^1_0dx\phi^s_{S}(x)=\int^1_0dx\phi^{\sigma}_{S}(x)=\frac{\bar{f}_{S}}{2\sqrt{6}},
\end{eqnarray}
and $\phi^T_{S}(x)=\frac{1}{6}\frac{d}{dx}\phi^{\sigma}_{S}(x)$. The
twist-2 LCDA can be expanded in the Gegenbauer polynomials:
\begin{eqnarray}
\phi_S(x,\mu)=\frac{1}{\sqrt{6}}\bar
f_S(\mu)6x(1-x)\sum_{m=1}^\infty B_m(\mu)C^{3/2}_m(2x-1).
\end{eqnarray}
The decay constants and the Gegenbauer moments of the twist-2 wave
function in two different scenarios  have been studied explicitly in
Refs.~\cite{Cheng:2005nb} using the QCD sum rule approach. As for
the explicit form of the Gegenbauer moments for the twist-3 wave
functions, there exist some uncertainties theoretically
\cite{Lu:2006fr}, thus we choice the asymptotic form for simplicity:
\begin{eqnarray}
\phi^s_S= \frac{1}{\sqrt {6}}\bar
f_f,\,\,\,\,\,\,\,\,\,\,\,\,\,\,\,\,\,\, \phi_S^T=\frac{1}{\sqrt
{6}}\bar f_S(1-2x).
\end{eqnarray}

For the vector mesons, the normalization for the twist-2 function
$\Phi_V$ and the twist-3 function $\Phi_v$ is given by
\begin{eqnarray}
 \int_0^1dx\Phi_V(x)=f_V, \qquad  \int^1_0 dx\Phi_v(x)=0,
\end{eqnarray}
where the definitions for $\Phi_v(x)$ can be found in
\cite{Beneke:1999br}. The general expressions of these LCDAs read
\begin{eqnarray}
 \Phi_V(x,\mu)=6x(1-x)f_V\left[1+\sum_{n=1}^\infty
 \alpha_n^V(\mu)C_n^{3/2}(2x-1)\right],
\end{eqnarray}
and
\begin{eqnarray}
 \Phi_v(x,\mu)=3f_V^\bot\left[2x-1+\sum_{n=1}^\infty
 \alpha_{n,\bot}^V(\mu)P_{n+1}(2x-1)\right],
\end{eqnarray}
where $P_n(x)$ are the Legendre polynomials.

In the calculation, the most important nonperturbative parameters are form
factors of $B\to S,V$ transitions, which are defined by \cite{BSW}:
\begin{eqnarray} \label{eq:FF}
   \la V(p')|V_\mu|B(p)\ra &=& -{\frac{1}{m_B+m_V}
}\,\epsilon_{\mu\nu\alpha \beta}\vp^{*\nu}P^\alpha
q^\beta  V^{BV}(q^2),   \non \\
 \la V(p')|A_\mu|B(p)\ra &=&  i\Big\{
(m_B+m_V)\vp^{*}_\mu A_1^{BV}(q^2)-\frac{\vp^{*}\cdot P}{m_B+m_V}\,P_\mu A_2^{BV}(q^2)\non\\
& &-2m_V\,\frac{\vp^{*}\cdot P}{q^2} \,q_\mu\big[A_3^{BV}(q^2)-A_0^{BV}(q^2)\big]\Big\}, \non\\
\la S(p')|A_\mu|B(p)\ra &=&
-i\Bigg[\left(P_\mu-\frac{m_B^2-m_S^2}{q^2}\,q_\mu\right)
F_1^{BS}(q^2)+ \frac{m_B^2-m_S^2}{q^2}q_\mu\,F_0^{BS}(q^2)\Bigg],
 \end{eqnarray}
with $P_\mu=(p+p')_\mu$, $q_\mu=(p-p')_\mu$.

To calculate the amplitudes, we start from the  effective
Hamiltonian responsible for $b\to s$ transitions, which  is given
by~\cite{Buchalla}
\begin{eqnarray}\label{eq:eff}
 {\cal H}_{\rm eff} &=& \frac{G_F}{\sqrt{2}} \biggl[V_{ub}
 V_{us}^* \left(C_1 O_1^u + C_2 O_2^u \right) + V_{cb} V_{cs}^* \left(C_1
 O_1^c + C_2 O_2^c \right) - V_{tb} V_{ts}^*\, \big(\sum_{i = 3}^{10}
 C_i O_i \big. \biggl. \nonumber\\
 && \biggl. \big. + C_{7\gamma} O_{7\gamma} + C_{8g} O_{8g}\big)\biggl] +
 {\rm h.c.}
\end{eqnarray}
In the above equation, $V_{qb} V_{qs}^*$~($q=u,c,t$) represent for
products of the Cabibbo-Kobayashi-Maskawa~(CKM) matrix elements,
$C_{i}$  are the responding Wilson coefficients, and $O_i$ are the
relevant four-quark operators whose explicit forms could be found,
for example, in Refs.~\cite{Buchalla}.

We now turn to study the short-distance contributions within the QCDF approach, where
the contribution of the nonperturbative sector is dominated by the
form factors and the nonfactorizable impact in the hadronic matrix
elements is controlled by hard gluon exchange. The hadronic matrix
elements of the decay can be written as
\begin{eqnarray}
 \langle M_1 M_2|O_i|B\rangle &=& \sum_{j}F_{j}^{B\rightarrow
M_1 }\int_{0}^{1}dx T_{ij}^{I}(x)\Phi_{M_1}(x)
\nonumber\\&+&\int_{0}^{1}d\xi\int_{0}^{1}dx\int_{0}^{1}dy
T_{i}^{II}(\xi,x,y)\Phi_{B}(\xi)\Phi_{M_1}(x)\Phi_{M_2}(y),
\end{eqnarray}
where $T_{ij}^{I}$ and $T_{i}^{II}$ denote short-distance  interactions and can be calculated perturbatively.
$\Phi_{X}(x)$ are the universal nonperturbative light-cone
distribution amplitudes. Using the weak effective Hamiltonian given
by Eq.(\ref{eq:eff}),  we then obtain the decay amplitudes as:
\begin{eqnarray}
A(B^- \to K^{*-}_0\phi ) &=&
i\frac{G_F}{\sqrt{2}}\sum_{p=u,c}\lambda_p^{(s)}
 \Bigg\{ \left( a_3+a_4^p+a_5-r_\chi^\phi(a_6^p-{\frac{1}{2}}a_8^p)-{\frac{1}{2}}(a_7+a_9+a_{10}^p)\right)_{ K^*_0\phi} \non \\
 &\times& 2f_\phi F_1^{BK^*_0}(m_\phi^2)m_Bp_c
 - f_Bf_\phi f_{K^*_0}\big(b_2\delta_u^p+b_3
 +b_{\rm 3,EW}\big)_{K^*_0\phi} \Bigg\},
\end{eqnarray}
\begin{eqnarray}
A(\ov B^0 \to \ov K^{*0}_0\phi ) &=&
i\frac{G_F}{\sqrt{2}}\sum_{p=u,c}\lambda_p^{(s)}
 \Bigg\{ \left( a_3+a_4^p+a_5-r_\chi^\phi(a_6^p-{\frac{1}{2}}a_8^p)-{\frac{1}{2}}(a_7+a_9+a_{10}^p)\right)_{ K^*_0\phi} \non \\
 &\times& 2f_\phi F_1^{BK^*_0}(m_\phi^2)m_Bp_c
 - f_Bf_\phi f_{K^*_0}\big(b_3
 -{\frac{1}{2}}b_{\rm 3,EW}\big)_{K^*_0\phi} \Bigg\},
\end{eqnarray}
\begin{eqnarray}
A(B^- \to \ov K^{*0}_0\rho^- ) &=&
i\frac{G_F}{\sqrt{2}}\sum_{p=u,c}\lambda_p^{(s)}
 \Bigg\{ -\left( a_4^p+r_\chi^{K^*_0}(a_6^p-{\frac{1}{2}}a_8^p)
 -{\frac{1}{2}}a_{10}^p\right)_{\rho K^*_0} \non \\
 &\times& 2f_{K_0^*}A_0^{B\rho}(m_{K_0^*}^2)m_Bp_c
 - f_Bf_\rho f_{K^*_0}\big(b_2\delta_u^p+b_3
 +b_{\rm 3,EW}\big)_{\rho K^*_0} \Bigg\}, \end{eqnarray}
\begin{eqnarray}
A(B^- \to K^{*-}_0\rho^0 ) &=&
i\frac{G_F}{2}\sum_{p=u,c}\lambda_p^{(s)}
 \Bigg\{-\left( a_1\delta_u^p+a_4^p+r_\chi^{K^*_0}(a_6^p+a_8^p)
 +a_{10}^p \right)_{\rho K^*_0} \non \\
 &\times& 2f_{K_0^*}A_0^{B\rho}(m_{K_0^*}^2)m_Bp_c+\left[a_2\delta_u^p+\frac{3}{2}
(a_9+a_7)\right]_{K^*_0\rho}2f_\rho F_1^{BK^*_0}(m_\rho^2)m_Bp_c
 \non \\
 &-& f_Bf_\rho f_{K^*_0}\big(b_2\delta_u^p+b_3
 +b_{\rm 3,EW}\big)_{\rho K^*_0} \Bigg\}, \end{eqnarray}
\begin{eqnarray}
 A(\ov B^0 \to K^{*-}_0\rho^+ ) &=&
i\frac{G_F}{\sqrt{2}}\sum_{p=u,c}\lambda_p^{(s)}
 \Bigg\{ -\left( a_1\delta_u^p+ a_4^p+r_\chi^{K^*_0}a_6^p
 +a_{10}^p+r_\chi^{K^*_0}a_8^p \right)_{\rho K^*_0} \non \\
 &\times&
 2f_{K_0^*}A_0^{B\rho}(m_{K_0^*}^2)m_Bp_c
 - f_Bf_\rho f_{K^*_0}\big(b_3
 -{\frac{1}{2}}b_{\rm 3,EW}\big)_{\rho K^*_0} \Bigg\}, \end{eqnarray}
\begin{eqnarray}
 A(\ov B^0 \to \ov K^{*0}_0\rho^0 ) &=&
i\frac{G_F}{2}\sum_{p=u,c}\lambda_p^{(s)}
 \Bigg\{ -\left( -a_4^p-r_\chi^{K^*_0}(a_6^p-{\frac{1}{2}}a_8^p)
 +{\frac{1}{2}}a_{10}^p \right)_{\rho K^*_0} \non \\
 &\times&
 2f_{K_0^*}A_0^{B\rho}(m_{K_0^*}^2)m_Bp_c+\left[a_2\delta_u^p+{\frac{3}{2}} (a_9+a_7)
\right]_{K^*_0\rho}2f_\rho F_1^{BK^*_0}(m_\rho^2)m_Bp_c
 \non \\
 &-& f_Bf_\rho f_{K^*_0}\big(-b_3
 +{\frac{1}{2}}b_{\rm 3,EW}\big)_{\rho K^*_0} \Bigg\}, \end{eqnarray}
\begin{eqnarray}
A(B^- \to K^{*-}_0\omega ) &=&
i\frac{G_F}{2}\sum_{p=u,c}\lambda_p^{(s)}
 \Bigg\{ \left[a_2\delta_u^p+2(a_3+a_5)+\frac{1}{2}(a_9+a_7)\right]_{K^*_0\omega}2f_\omega
F_1^{BK^*_0}(m_\omega^2)m_Bp_c
 \non \\
  &-& \left( a_1\delta_u^p+a_4^p+r_\chi^{K^*_0}(a_6^p+a_8^p)
 +a_{10}^p \right)_{\omega K^*_0}
 2f_{K_0^*}A_0^{B\omega}(m_{K_0^*}^2)m_Bp_c \non \\
 &-& f_Bf_\omega f_{K^*_0}\big(b_2\delta_u^p+b_3
 +b_{\rm 3,EW}\big)_{\omega K^*_0} \Bigg\}, \end{eqnarray}
\begin{eqnarray}
 A(\ov B^0 \to \ov K^{*0}_0\omega ) &=&
i\frac{G_F}{2}\sum_{p=u,c}\lambda_p^{(s)}
 \Bigg\{ \left[a_2\delta_u^p+2(a_3+a_5)+\frac{1}{2}
(a_9+a_7)\right]_{K^*_0\omega}2f_\omega F_1^{BK^*_0}(m_\omega^2)m_Bp_c
 \non \\
 &-& \left(a_4^p+r_\chi^{K^*_0}(a_6^p-{\frac{1}{2}}a_8^p)
 -{\frac{1}{2}}a_{10}^p \right)_{\omega K^*_0}
 2f_{K_0^*}A_0^{B\rho}(m_{K_0^*}^2)m_Bp_c \non \\
 &-& f_Bf_\omega f_{K^*_0}\big(b_3
 -{\frac{1}{2}}b_{\rm 3,EW}\big)_{\omega K^*_0} \Bigg\};
\end{eqnarray}
where the ratios $r_\chi^V$ and $r_\chi^S$ are defined as
\begin{eqnarray} \label{eq:rchiS}
 r_\chi^V(\mu)=\frac{2m_V}{m_b(\mu)}\,\frac{f_V^\bot(\mu)}{f_V} , \qquad\quad
 r_\chi^S(\mu)=\frac{2m_S^2}{m_b(\mu)(m_2(\mu)-m_1(\mu))}.
\end{eqnarray}
The order of the arguments of the $a_i^p(M_1M_2)$ and $b_i(M_1M_2)$
coefficients is dictated by the subscript $M_1M_2$, where $M_1$
shares the same spectator quark with the $B$ meson and $M_2$ is the
emitted meson. For the annihilation part, $M_1$ is referred to the
one containing an anti-quark from the weak vertex, and $M_2$
contains a quark from the weak vertex. Combining the short-distance
nonfactorizable corrections, the effective Wilson coefficients
$a_i^p$ have the expressions
 \begin{eqnarray} \label{eq:ai}
 a_i^p(M_1M_2) &=& \left(C_i+\frac{C_{i\pm1}}{N_c}\right)N_i(M_2)
  +\frac{C_{i\pm1}}{N_c}\,\frac{C_F\alpha_s}{4\pi}\Big[V_i(M_2)+
\frac{4\pi^2}{N_c}H_i(M_1M_2)\Big]+P_i^p(M_2),
\end{eqnarray}
where $V_i(M_2)$ account for vertex corrections, $H_i(M_1M_2)$ for
hard spectator interactions  and $P_i(M_2)$ for penguin
contractions. The coefficients $b_i$ and $b_{i,{\rm EW}}$ stand for
the contribution of annihilation diagrams.

In QCDF approach, the end-point singularities appear in calculating
the twist-3 spectator and annihilation amplitudes. Since the
treatment of endpoint divergences is model dependent, subleading
power corrections generally can be studied only in a
phenomenological way. As the most popular way, the end-point
divergent integrals are treated as signs of infrared sensitive
contributions and   parameterized by \cite{Beneke:1999br}:
\begin{equation}\label{treat-for-anni}
\int_0^1 \frac{\!dy}{y}\, \to X_A =(1+\rho_A e^{i\phi_A}) \ln
\frac{m_B}{\Lambda_h} ,
 \end{equation}
with the unknown real parameters $\rho_A$ and $\phi_A$. More
discussion about them will be in Section.\ref{sec:4}.

\section{The Family Nonuniversal $Z^\prime$ Model}\label{sec:3}
In this section, we will review the main part of the family
nonuniversal $Z^\prime$ model briefly. In the current work, for
simplicity, we only focus on the models in which the interactions
between the $Z^\prime$ boson and fermions are flavor nonuniversal
for left-handed couplings and flavor diagonal for right-handed
cases. Of course, the analysis can be straightly extended to general
cases in which the right-handed couplings are also nonuniversal
across generations. The basic formulas of the $Z^\prime$ model with
family nonuniversal and/or nondiagonal couplings have been presented
in Refs.\cite{Langacker:2008yv, Langacker:2000ju}, to which we refer
readers for detail. Here, we just review the ingredients needed in
this work.

In the gauge basis, the neutral current Lagrangian induced by the
$Z^\prime$ boson  can be written as
\begin{eqnarray}
\label{eq:Zpr} {\cal L}^{Z^\prime} =-g_2J^\prime_{\mu} Z^{\prime
\mu} ~,
\end{eqnarray}
where $g_2$ is the gauge coupling associated with the additional
$U(1)^\prime$ group at the $M_W$ scale.  Neglecting the
renormalization group (RG) running effect between $M_W$ and $M_{Z'}$
and the mixing between $Z^\prime$ and $Z$ boson of SM, we present
the chiral current as
\begin{eqnarray}
J'_{\mu} = \sum_{i,j} {\overline \psi_i^I} \gamma_{\mu}
  \left[ (\epsilon_{\psi_L})_{ij} P_L + (\epsilon_{\psi_R})_{ij} P_R \right]
  \psi^I_j ~,
\end{eqnarray}
where the sum extends over the flavors of fermions, the chirality
projection operators are $P_{L,R} \equiv (1 \mp \gamma_5) / 2$, the
superscript $I$ stands for the weak interaction eigenstates, and
$\epsilon_{\psi_L}$ ($\epsilon_{\psi_R}$) denote the left-handed
(right-handed) chiral couplings. $\epsilon_{\psi_L}$ and
$\epsilon_{\psi_R}$ are required to be  hermitian so as to arrive a
real Lagrangian. Accordingly, the mass eigenstates of the chiral
fields can be defined by $\psi_{L,R} = V_{\psi_{L,R}} \psi_{L,R}^I$,
and the usual CKM matrix is given by $V_{\rm CKM} = V_{u_L}
V_{d_L}^{\dagger}$. Then, the chiral $Z^\prime$ coupling matrices in
the physical basis of up-type and down-type quarks are,
respectively,
\begin{eqnarray}
B^X_u \equiv V_{u_X} \epsilon_{u_X} V_{u_X}^{\dagger} ~, ~~ B^X_d
\equiv V_{d_X} \epsilon_{d_X} V_{d_X}^{\dagger} ~~ (X = L,R).
\end{eqnarray}
If the $\epsilon$ matrices are not proportional to the identity, the
$B$ matrices will have non-zero off-diagonal elements, which  induce
FCNC interactions at the tree level directly. In this work, we
assume that the right-handed couplings are diagonal for simplicity.
Thereby, the effective Hamiltonian of the ${\bar b} \to {\bar s} q
{\bar q}(q=u,d)$ transitions mediated by the $Z^\prime$ is
\begin{eqnarray}
{\cal H}_{\rm eff}^{Z'}=\frac{2 G_F}{\sqrt{2}} \left(\frac{g_2
M_Z}{g_1 M_{Z'}}\right)^2 B^{L*}_{sb}
   ({\bar b}s)_{V-A} \sum_q \left( B^L_{qq} ({\bar q}q)_{V-A}   + B^R_{qq} ({\bar q}q)_{V+A} \right) + \mbox{h.c.} ~,
\label{eqn:Heff1} \end{eqnarray}
where $g_1=e/(\sin{\theta_W}\cos{\theta_W})$ and $M_{Z^{\prime}}$
the mass of the new gauge boson. We note the above operators of the
forms $({\bar b}s)_{V-A} ({\bar q}q)_{V-A}$ and $({\bar b}s)_{V-A}
({\bar q}q)_{V+A}$ already exist in SM, so that we  represent the
$Z^\prime$ effect as a modification to the Wilson coefficients of
the corresponding operators. Hence, we rewrite the
eq.(\ref{eqn:Heff1}) as
\begin{eqnarray}
{\cal H}_{\rm eff}^{Z'} &= &- \frac{G_F}{\sqrt{2}} V_{tb}^* V_{ts}
\sum_q \left( \Delta C_3 O_3^{(q)} +
  \Delta C_5 O_5^{(q)} + \Delta C_7 O_7^{(q)} + \Delta C_9 O_9^{(q)} \right) +
\mbox{h.c.}, \label{eqn:Heff2}
\end{eqnarray}
where the additional contributions to the SM Wilson coefficients at
the $M_W$ scale in terms of $Z'$ parameters are given by
\begin{eqnarray} && \Delta C_{3(5)} = - \frac{2}{3 V_{tb}^* V_{ts}}
\left(\frac{g_2 M_Z}{g_1
    M_{Z'}}\right)^2 B^{L*}_{sb} \left(B^{L(R)}_{uu} + 2 B^{L(R)}_{dd} \right)
\\
&& \Delta C_{9(7)} = -\frac{4}{3 V_{tb}^* V_{ts}} \left(\frac{g_2
M_Z}{g_1
    M_{Z'}}\right)^2 B^{L*}_{sb} \left(B^{L(R)}_{uu} - B^{L(R)}_{dd} \right).
\end{eqnarray}
Thus we can have a $Z'$ contribution to the QCD penguins $\Delta
C_{3(5)}$ as well as the EW penguins $\Delta C_{9(7)}$, in the light
of the results found by Buras et al. \cite{Buchalla}. In order to
show that the new physics is primarily manifest in the EW penguins,
we assume $B^{L(R)}_{uu} \simeq -2 B^{L(R)}_{dd}$, which have been
used widely
\cite{Barger:2009hn,Cheung:2006tm,Chang:2009wt,Arhrib:2006sg}. As a
result, the $Z'$ contributions to the Wilson coefficients at the
weak scale are
\begin{eqnarray}
&&\Delta C_{3(5)}= 0 ~, \\
&&\Delta C_{9(7)} = 4 \frac{|V_{tb}^* V_{ts}|}{V_{tb}^* V_{ts}}
\xi^{LL(R)} e^{- i \phi_L} ~, \label{eqn:c97} \end{eqnarray}
where
\begin{eqnarray} \xi^{LX} &\equiv& \left(\frac{g_2 M_Z}{g_1 M_{Z'}}\right)^2
\left|\frac{B^{L*}_{sb} B^X_{dd}}{V_{tb}^* V_{ts}}\right| ~~
(X=L,R) ~, \label{eqn:xi} \\
\phi_L &\equiv& {\rm Arg}[B^L_{sb}] ~. \end{eqnarray}
Because of the hermiticity of the effective Hamiltonian, the
diagonal elements of the effective coupling matrix must be real.
However, the off-diagonal elements, such as $B^L_{sb}$, generally
may contain new weak phases. Moreover, the relation $B^{L(R)}_{ss}
\simeq B^{L(R)}_{dd}$ follows from the assumptions of universality
for the first two families, as required by $K$ and $\mu$ decay
constraints \cite{Langacker:2000ju}. Since the major objective of
our work is searching for new physics signal, rather than producing
acute numerical results, we also assume $B^{L}_{qq} \simeq
B^{R}_{qq}$, because  we expect that $|B^L_{qq}|$ and $|B^R_{qq}|$
should have the same order of magnitude.

It should be emphasized that the other SM Wilson coefficients may
also receive contributions from the $Z^{\prime}$ boson through
renormalization group~(RG) evolution. With our assumption that no
significant RG running effect between $M_Z^{\prime}$ and $M_W$
scales, the RG evolution of the modified Wilson coefficients is
exactly the same as the ones in SM~\cite{Buchalla}. The numerical
results of Wilson coefficients in the naive dimensional
regularization~(NDR) scheme at the scale $\mu=2.1 {\rm
GeV}$~($\mu_h=1 {\rm GeV} $) are listed in Table~\ref{Wilson} for
convenience.

In summary, we list here our simplifications to a general $Z^\prime$
model: we assume (i) no right-handed flavor-changing couplings
($B^R_{ij} = 0$ for $i \neq j$), (ii) no significant RG running
effect between $M_{Z^\prime}$ and $M_{W}$ scales, (iii) negligible
$Z^\prime$ effect on the QCD penguin ($\Delta C_{3,5} = 0$) so that
the new physics is manifestly isospin violating, (iv) $|B^L_{qq}|$
and $|B^R_{qq}|$ are same so as to reduce the number of parameters.
With these simplifications, we have only two parameters left in the
model. So, this approach provides a minimal way to introduce the
$Z^\prime$ effect in the concerned decay modes. Of course, more
general $Z^\prime$ models are possible.

Now, the only task left is to constraint the parameters within the
existing experimental data. Generally, $g_2/g_1 \sim 1$ is expected,
if both the $U(1)$ gauge groups have the same origin from some grand
unified theories. We also hope $ M_Z/M_{Z^\prime}\sim 0.1$ so that TeV
scale neutral $Z^\prime$ boson could be detected at LHC.
Theoretically, one can fit the left three parameters $|B^{L}_{sb}|$,
$|B^X_{dd}|$ and new weak phase $\phi_{L}$ with the accurate data
from $B$ factories and other experiments such as Tavatron and LHC.
For example,  $B^L_{sb}$ and $\phi_{L}$ could be extracted from
$B_s$-$\bar B_s$ mixing as well as $B \to K^{(*)} \ell^+\ell^-$
decays. To resolve the mass difference between $B_s$ and $\bar B_s$,
$|B^{L}_{sb}|\sim|V_{tb}V_{ts}^{*}|$ is required
\cite{Barger:2009hn,Chang:2009wt,Alok:2010ij}. In
Refs.\cite{Chang:2009wt}, the authors  got the $\phi_{L}$ is about
$-80^\circ$ by fitting data of $B_s-\bar{B_s}$ mixing and $B \to
K^{(\ast)} l^+ l^-$ decays. Subsequently, with $B^L_{sb}$ and
$\phi_L$ arrived and experimental data of $B \to \pi \pi, K \pi,
K\rho$ and $K^{(*)}\phi$, $B^L_{qq}$ and $B^R_{qq}$ could be
extracted analogously. Specifically, the $CP$ asymmetries in $B \to
K\phi, K \pi$ can be resolved if $|B^L_{sb}B^{L,R}_{ss} |\sim
|V_{tb}V^{*}_{ts} |$, which indicates $|B^{L,R}_{qq}| \sim 1$.
However, we have one remark here. In dealing with the nonleptonic
$B$ decays, because different groups used different factorization
approach, the fitted results are different, but all results have
same order. Noted that the detailed constraint of these parameters is beyond the 
scope of current work and can be found in many references \cite{Cheung:2006tm,Chang:2009wt}.
Summing up above analysis, we thereby
assume that $\xi=\xi^{LL}=\xi^{LR} \in (10^{-3},10^{-2})$ and
$\phi_L \in (-60^\circ, -90^\circ)$ so as to prob the new physics
effect for maximum range.

\begin{table}[t]
 \begin{center}
 \caption{The Wilson coefficients $C_i$ within SM and with the contribution
 from $Z^{\prime}$ boson included in NDR scheme at the scale $\mu=2.1~~\rm{GeV}$ and
 $\mu_h=1.0~~\rm{GeV}$.}
 \label{Wilson}
 \vspace{0.1cm}
 \small
 \doublerulesep 0.7pt \tabcolsep 0.04in
 \begin{tabular}{c|cc|cc}\hline\hline
 Wilson                   &\multicolumn{2}{c|}{$\mu=2.1~~\rm{GeV}$}
 &\multicolumn{2}{c}{$\mu_h=1.0~~\rm{GeV}$} \\\cline{2-3}\cline{4-5}
 coefficients             &$C_i^{SM}$ &$\Delta C_i^{Z^{\prime}}$
 &$C_i^{SM}$ &$\Delta C_i^{Z^{\prime}}$\\ \hline\hline
 $C_1$                    &$1.135$    &$0$                                   &$1.224$    &$0$\\
 $C_2$                    &$-0.283$   &$0$                                   &$-0.429$   &$0$ \\
 $C_3$                    &$0.021$    &$0.09\xi^{LL}-0.02\xi^{LR}$           &$0.034$    &$0.15\xi^{LL}-0.04\xi^{LR}$\\
 $C_4$                    &$-0.049$   &$-0.20\xi^{LL}+0.01\xi^{LR}$          &$-0.072$   &$-0.31\xi^{LL}+0.03\xi^{LR}$\\
 $C_5$                    &$0.010$    &$0.03\xi^{LL}+0.02\xi^{LR}$           &$0.010$    &$0.02\xi^{LL}+0.02\xi^{LR}$\\
 $C_6$                    &$-0.06$    &$-0.26\xi^{LL}+0.03\xi^{LR}$          &$-0.104$   &$-0.44\xi^{LL}+0.07\xi^{LR}$\\
 $C_7/{\alpha}_{em}$      &$-0.018$   &$5.3\xi^{LL}-461\xi^{LR}$             &$-0.023$   &$6.3\xi^{LL}-457\xi^{LR}$\\
 $C_8/{\alpha}_{em}$      &$0.081$    &$2.43\xi^{LL}-286\xi^{LR}$            &$0.134$    &$4.8\xi^{LL}-497\xi^{LR}$\\
 $C_9/{\alpha}_{em}$      &$-1.266$   &$-594\xi^{LL}+6.1\xi^{LR}$            &$-1.366$   &$-643\xi^{LL}+7.8\xi^{LR}$\\
 $C_{10}/{\alpha}_{em}$   &$0.321$    &$178\xi^{LL}-1.0\xi^{LR}$             &$0.483$    &$257\xi^{LL}-1.9\xi^{LR}$\\
 $C_{7{\gamma}}$          &$-0.345$   &---                                         &$-0.395$    &---\\
 $C_{8g}$                 &$-0.161$   &---                                         &$-0.181$   &---\\
 \hline \hline
 \end{tabular}
 \end{center}
 \end{table}
\section{Numerical Results and Discussion}\label{sec:4}
In this section, to begin with, we will give the parameters used in this work. Since
it is not clear whether the scalar meson $K_0^*(1430)$ belongs to
the first orbital excited state (S1) or the low lying resonance
(S2), we will calculate the processes under both scenarios. In the
calculation, the decay constants and Gegenbauer moments obtained
within the QCD sum rules method under different scenarios are
presented as follows \cite{Cheng:2005nb}:
\begin{eqnarray}
 &\mathbf{ S1}:& \bar f_{K_0^*}(1.0 \mathrm{GeV})=-300  \mathrm{MeV};
       \bar f_{K_0^*}(2.1 \mathrm{GeV})=-370 \mathrm{MeV};
      B_1(1.0\mathrm{GeV})=0.58;\non\\
 &&
       B_1(2.1 \mathrm{GeV})=0.39;     B_3(1.0 \mathrm{GeV})=-1.20;
      B_3(2.1\mathrm{GeV})=-0.70; \\
 &\mathbf{ S2} :& \bar f_{K_0^*}(1.0\mathrm{GeV})=445 \mathrm{MeV};
     \bar f_{K_0^*}(2.1\mathrm{GeV})=550\mathrm{MeV};
      B_1(1.0\mathrm{GeV})=-0.57;\non\\
&&
      B_1(2.1\mathrm{GeV})=-0.39;    B_3(1.0\mathrm{GeV})=-0.42;
      B_3(2.1\mathrm{GeV})=-0.25.
\end{eqnarray}
In QCD sum rules method, the major parameter is the Borel window,
which takes large uncertainty to the parameters listed above. In
Ref.\cite{Cheng:2005nb}, the authors had discussed the errors caused
by them in great detail and found that $B_{1,3}$ will take $30\%$
changes. As a result, we will not discuss this part any more in the
current work.

For the vector mesons, the longitudinal and transverse decay
constants  are list as:
\begin{eqnarray}
 && f_\rho=216~~\mathrm{MeV}, \qquad f_\omega=187~~\mathrm{MeV},
 \qquad~~ f_\phi=215~~\mathrm{MeV}, \non \\
 && f_\rho^\bot=165~~\mathrm{MeV}, \qquad f_\omega^\bot=151~~\mathrm{MeV},  \qquad
 f_\phi^\bot=186~~\mathrm{MeV}\,,
\end{eqnarray}
where the values are taken from \cite{Ball:2006eu}. In the LCADs of
vectors, the Gegenbauer moments $\alpha_n^V$ and $\alpha^V_{n,\bot}$
have been studied within the QCD sum rule method. Here, we will
employ the most recent updated values \cite{Ball:2007rt}
\begin{eqnarray}
  \alpha_2^{\rho,\omega}=0.15,\qquad   \alpha_{2,\bot}^{\rho,\omega}=0.14,
  \qquad   \alpha_2^\phi=0.18,\qquad   \alpha_{2,\bot}^\phi=0.14,
\end{eqnarray}
and $\alpha_1^V=0$, $\alpha_{1,\bot}^V=0$.

As stated earlier, various form factors for $B\to S,V$
transitions have been evaluated in cLFQM \cite{Cheng:2003sm}. In
this model form factors are first calculated in the spacelike region
and their momentum dependence is fitted to a 3-parameter form
\begin{eqnarray} \label{eq:FFpara}
 F(q^2)=\frac{F(0)}{1-a(q^2/m_{B}^2)+b(q^2/m_{B}^2)^2}.
\end{eqnarray}
The parameters $a$, $b$ and $F(0)$ relevant for our purposes are
summarized in Table.\ref{tab:LFBtopi}.
\begin{table}[b]
\caption{Form factors of $B\to \rho, K^*_0(1430)$ transitions
obtained in the covariant light-front model \cite{Cheng:2003sm}.}
\label{tab:LFBtopi}
\begin{tabular}{| c c c c c || c c c c c |}
\hline \hline ~~~$F$~~~~~
    & $F(0)$~~~~~
    & $F(q^2_{\rm max})$~~~~
    &$a$~~~~~
    & $b$~~~~~~
& ~~~ $F$~~~~~
    & $F(0)$~~~~~
    & $F(q^2_{\rm max})$~~~~~
    & $a$~~~~~
    & $b$~~~~~~
 \\
    \hline
$V^{B\rho}$
    & $0.27$
    & $0.79$
    & 1.84
    & 1.28
&$A^{B\rho}_0$
    & 0.28
    & 0.76
    & 1.73
    & 1.20
    \\
$A^{B\rho}_1$
    & 0.22
    & 0.53
    & 0.95
    & 0.21
&$A^{B\rho}_2$
    & $0.20$
    & $0.57$
    & 1.65
    & 1.05
    \\
$F^{BK^*_0}_1$[S1]
    & $0.21$
    & $0.52$
    & 1.59
    & 0.91
&$F^{BK^*_0}_0$[S1]
    & 0.21
    & 0.30
    & 0.59
    & 0.09  
    \\
$F^{BK^*_0}_1$[S2]
    & $0.26$
    & $0.70$
    & 1.52
    & 0.64
&$F^{BK^*_0}_0$[S2]
    & 0.26
    & 0.33
    & 0.44
    & 0.05\\
    \hline\hline
\end{tabular}
\end{table}

In Refs.\cite{Cheng:2005nb, Cheng:2007st,Li:2011kw}, it was
found that in decay modes with scalars the main theoretical
uncertainties are due to the weak annihilations, especially for the
penguin dominated ones. In $B \to PP, PV$ decays, the annihilation
amplitudes are helicity suppressed because the helicity of one of
final states cannot match with that of its quarks. However, this
helicity suppression can be alleviated in the decay modes with
scalar because of nonvanishing orbital angular momentum. Thus, weak
annihilation contribution to $B \to SP(V)$ is much larger than the
$B \to PP(V)$ case. However, as stated before, the end-point
singularity appears in calculating the annihilation contribution,
and then two free parameters, $\rho_A$ and $\phi_A$, are introduced
phenomenally. In Ref. \cite{Cheng:2007st}, it is found that the
behavior of $SV$ is similar to the longitudinal part of $VV$.
Fortunately, with experimental data, it presents the moderate value
of nonuniversal annihilation phase $\phi_A=-40^{\circ}$ for $B \to
VV$ decay modes \cite{Beneke:1999br}. Therefor, for $B \to SV$, we
conservatively take $\phi_A=(-40\pm20)^{\circ}$ with $\rho_A=0.6\pm
0.2$, which also assures that the hadronic uncertainties are
considerably reduced. Furthermore, the endpoint divergence $X_H$ in
the hard spectator contributions can also be parameterized in the
same manner.

Within above parameters and formulas, we calculate the branching
fractions of these decays in SM and the family nonuniversal
$Z^\prime$ model under two different scenarios. Together with
partial experimental results, the results under are
exhibited in Table.\ref{Table:3}, respectively. For the center
values, we adopt $\xi=0.005$ and $\phi_L^{sb}=-80^\circ$. For all
theoretical predictions, the first errors arise from the power
corrections of weak annihilation and hard spectator interactions
characterized by the parameters $X_{A,H}$. To obtain the second
errors of the $Z^\prime$ model results, we scan randomly the points
in their own possible parameter spaces.

\begin{table}[t]
\begin{center}
\caption{Branching fractions (in units of $10^{-6}$) under the different Scenarios}
\label{Table:3}
\begin{tabular}{c|cc|cc|c }
\hline\hline
&\multicolumn{2}{|c|}{S1}& \multicolumn{2}{|c|}{S2}& \\
Decay Mode  &SM & SM+$Z^\prime$& SM & SM+$Z^\prime$ & Expt \\
\hline $B^-\to K_0^{*-}\phi$ &$2.4^{+5.1}_{-1.8}$
&$3.8^{+5.6+1.8}_{-2.5-1.2}$ &$22.6^{+19.7}_{-8.6}$
&$16.5^{+18.5+5.0}_{-10.3-7.9}$
&$7.0\pm 1.3\pm 0.9$\\

$\overline B^0\to K_0^{*0}\phi$ &$2.2^{+4.9}_{-1.7}$
&$4.7^{+5.2+4.4}_{-2.6-2.2}$ &$22.4^{+19.4}_{-8.4}$
&$21.2^{+18.9+3.0}_{-8.4-4.6}$
&$4.3\pm 0.6 \pm 0.4$\\

$B^-\to \bar K_0^{*0}\rho^-$ &$11.7^{+8.4}_{-4.0}$
&$11.5^{+7.7+3.6}_{-4.7-3.8}$ &$45.5^{+20.6}_{-10.4}$
&$41.4^{+19.7+4.4}_{-11.9-5.7}$
& \\

$B^-\to \bar K_0^{*-}\rho^0$ &$7.2^{+4.5}_{-2.3}$
&$18.2^{+5.9+27.4}_{-3.4-10.0}$ &$17.6^{+9.0}_{-4.4}$
&$15.9^{+8.4+6.9}_{-6.1-5.6}$
& \\

$\bar B^0\to \bar K_0^{*0}\rho^0$ &$4.6^{+1.7}_{-0.7}$
&$3.9^{+1.4+6.4}_{-0.8-1.8}$ &$24.5^{+7.0}_{-3.8}$
&$33.3^{+7.7+32.1}_{-4.8-8.5}$
&$27\pm 5.5$\\

$\bar B^0\to \bar K_0^{*-}\rho^+$ &$10.7^{+8.5}_{-3.7}$
&$14.4^{+9.2+8.7}_{-5.1-4.1}$ &$44.7^{+21.3}_{-10.5}$
&$54.1^{+22.4+25.7}_{-12.6-8.6}$
&$28\pm12$\\

$B^-\to \bar K_0^{*-}\omega$ &$3.6^{+3.1}_{-1.4}$
&$7.8^{+3.7+10.1}_{-1.7-3.8}$ &$12.6^{+7.2}_{-3.5}$
&$13.7^{+7.5+4.8}_{-3.9-1.4}$
&\\

$\bar B^0\to \bar K_0^{*0}\omega$ &$3.9^{+1.3}_{-0.6}$
&$4.0^{+1.3+1.2}_{-0.7-0.6}$ &$10.6^{+4.4}_{-2.1}$
&$10.7^{+4.4+5.4}_{-2.7-2.6}$
&\\
\hline \hline
\end{tabular}
\end{center}
\end{table}
Comparing our predictions of SM with those in Ref.
\cite{Chen:2007qj} (considering the typos), there are few
differences. Some reasons are list as follows: (1) In the Ref.
\cite{Chen:2007qj}, for the parameterizations of singularities, the
center values correspond to $\rho_{A,H}=0$ and $\phi_{A,H}=0$, while
we set $\rho_{A,H}=0.6$ and $\phi_{A,H}=-40^\circ$; (2) The
difference of Wilson coefficients, caused by the top quark mass and
other part parameters, will change the results slightly; (3) In this
work, the different form factors of $B \to K_0^*(1430)$ are used
under different scenarios, but they adopted same values in the
ref.\cite{Cheng:2007st}.

In Tables.\ref{Table:3}, for $K_0^*(1430)\phi$ channels, though the
central values of the predicted under S1 are smaller than the
experimental data, they are accommodated with the large
uncertainties. However, in S2, the theoretical results are much
larger than the data, and cannot agree with data even with
uncertainties. It should be noted that in this work we have not
included the errors from uncertainties of meson distribution
amplitudes ($B_1$ and $B_3$). Even with those uncertainties, the
theoretical results are still larger than the upper limits of the
data. These theoretical results also agree with the results from
pQCD approach \cite{Kim:2009dg}. For $\bar B^0\to \bar
K_0^{*0}\rho^0$ channels, contrary to $K_0^*(1430)\phi$, the result
of S2 agree with data well and the prediction of S1 is much smaller
than the data. Since for $\bar B^0\to \bar K_0^{*-}\rho^+$ there is
large uncertainty in the experimental data, the theoretical results
under both S1 and S2 can accommodate the data with large
uncertainties theoretically. That's to say, it is impossible to
explain all data under one settled scenario simultaneously.

When adding the contribution of the $Z^\prime$ gauge boson, as shown
in the table, the $Z^\prime$ gauge boson  changes the branching
fractions under both two different scenarios. For $B \to
K_0^*\phi$ channels dominated by the weak annihilation, the
$Z^\prime$ will enhance the branching fractions in S1, while in S2
the branching ratios are decreased. The reason is that the weak
annihilation is proportional to the decay constant $f_{K_0^*}$,
which has different sign in different scenarios. For $B^-\to \bar
K_0^{*0}\rho^-$ and $\bar B^0\to \bar K_0^{*-}\rho^+$, as the scalar
particle is the emitted particle, the whole amplitudes are
proportional to the decay constant $f_{K_0^*}$, thus the new physics
contribution have same behavior in different scenarios. For channels
with $\rho^0 $ or $\omega$, the spectator quarks enters not only the
scalars but also the vectors, the amplitudes become more complicate, and
we cannot describe the relation between new physics and branching
fractions apparently.

Compared to the experimental data, the $Z^\prime$ boson could change
the branching fractions remarkably and alleviate the disparities.
However, we cannot achieve a definite conclusion yet whether $
K_0^*$ belongs to the ground states or the first orbital
excited states. Moreover, for most modes except $B^-\to \bar
K_0^{*0-}\rho^0(\omega)$,  the new physics contribution might be
clouded by the uncertainties taken by the weak annihilations. Thus,
it is also very difficult to search for $Z^\prime$ effect in these
decays. Specifically, for decays $B^-\to \bar
K_0^{*-}\rho^0(\omega)$, $Z^\prime$ boson could enhance the
branching fractions more than 2 times,  we hope these two channels
could be measured in the LHC or Super-b factories in future so as to probe the 
$Z^\prime$ gauge boson.

To test the isospin symmetry and prob new physics, we define two
ratios:
\begin{eqnarray}
 R_1&=&\frac{Br(\bar B^0\to \bar K_0^{*0}\rho^0)}{Br(\bar B^0\to \bar
 K_0^{*-}\rho^+)}=0.96^{+1.07}_{-0.43}~~~[\mathrm{Exp.}]; \\
 R_2&=&\frac{\tau(\bar B^0)}{\tau(B^-)}\cdot\frac{Br( B^-\to \bar K_0^{*-}\phi)}{Br(\bar B^0\to \bar
 K_0^{*0}\phi)}=1.52^{+0.71}_{-0.51}~~~[\mathrm{Exp.}],
\end{eqnarray}
where the experimental results are also given and all uncertainties
are added in quadrature. In the isospin limit, $R_1=1/2$ and $R_2=1$
are expected to hold. Here, we list the theoretical results under
different scenarios in different models:
\begin{eqnarray}
R_1[\mathrm{SM}]=\left\{
      \begin{array}{ll}
        0.43^{+0.13}_{-0.10}, & \hbox{S1;} \\
        0.55^{+0.08}_{-0.08}, & \hbox{S2.}
      \end{array}
    \right.;\;\;\;\;\;\;\;\;\;\;\;\;\;\;\;\;
R_1[\mathrm{SM}+Z^\prime]=\left\{
      \begin{array}{ll}
        0.27^{+0.15+0.31}_{-0.08-0.17}, & \hbox{S1;} \\
        0.62^{+0.09+0.20}_{-0.09-0.07}, & \hbox{S2.}
      \end{array}
    \right.
\end{eqnarray}
\begin{eqnarray}
R_2[\mathrm{SM}]=\left\{
      \begin{array}{ll}
        1.00^{+0.04}_{-0.04}, & \hbox{S1;} \\
        0.94^{+0.00}_{-0.01}, & \hbox{S2.}
      \end{array}
    \right.;\;\;\;\;\;\;\;\;\;\;\;\;\;\;\;\;
R_2[\mathrm{SM}+Z^\prime]=\left\{
      \begin{array}{ll}
        0.76^{+0.16+0.24}_{-0.18-0.18}, & \hbox{S1;} \\
        0.73^{+0.16+0.19}_{-0.27-0.24}, & \hbox{S2.}
      \end{array}
    \right.;
\end{eqnarray}
In the above results, the theoretical uncertainties are reduced
since they are ratios of branch fractions.   We  see that the symmetries are almost held in SM. 
However, the data shows that the isospin symmetries are violated, which means that the large
weak annihilation may break the isospin symmetry remarkably. When
adding $Z^\prime$ contribution, except R1 under S2, the isospin
symmetries are broken in an opposite direction. However, the family
nonuniversal $Z^\prime$ model cannot be ruled out due to large
uncertainties in the experiments.

Finally, we will discuss the  $CP$ asymmetries of these decays. For
the charged mode $B^-\to K_0^{*-}\phi$, because $|V_{ub}V_{us}|
(\lambda^4) \ll|V_{tb}V_{ts}| (\lambda^2)$ and there is no tree
contribution in  the neutral mode $\overline B^0\to K_0^{*0}\phi$,
the direct $CP$ asymmetries are almost zero in both SM and the
$Z^\prime$ model. For $B\to K_0^{*}\rho$, although the CKM elements
are suppressed, the tree operators with large Wilson coefficients
appear in the emission diagrams, so the amplitudes of tree and
penguin may have comparable magnitudes. Thus, large $CP$ asymmetries
in these decays are expected, just like decays $B\to K \pi$ and
$B\to K \rho$. In Table.\ref{Table:4}, we give the $CP$ asymmetries
of $B\to K_0^{*}\rho$ in both SM and the concerned new physics model
under different scenarios. From the table, we firstly note that
$\bar B^0\to \bar K_0^{*0}\rho(\omega)$ have large asymmetries, and
different scenarios have different signs but with large
uncertainties. If we can calculate the annihilation accurately
within some effective approach in future, this parameter could be used to
distinguish the scenarios. Secondly, for $B^-\to \bar K_0^{*-}
\rho^0(\omega)$, the $Z^\prime$ could change the signs of the center
values, and these two decays can be used in probing new physics
effect.

\begin{table}[t]
\begin{center}
\caption{The direct $CP$ asymmetry ($\%$) under the different
scenarios} \label{Table:4}
\begin{tabular}{c|cc|cc}
\hline\hline
&\multicolumn{2}{|c|}{S1}& \multicolumn{2}{|c}{S2} \\
Decay Mode  &~~~SM~~~ & ~~~SM+$Z^\prime$~~~& ~~~SM~~~ & ~~~SM+$Z^\prime$ ~~~ \\
\hline
 $B^-\to \bar K_0^{*0}\rho^-$ &$6^{+4}_{-2}$
&$6^{+6+1}_{-3-1}$ &$2^{+2}_{-1}$ &$2^{+2+0}_{-1-0}$
\\

$B^-\to \bar K_0^{*-}\rho^0$ &$4^{+4}_{-3}$ &$-3^{+2+4}_{-1-2}$
&$-1^{+3}_{-4}$ &$6^{+3+10}_{-3-6}$
\\

$\bar B^0\to \bar K_0^{*0}\rho^0$ &$9^{+26}_{-38}$
&$24^{+52+21}_{-42-12}$ &$-11^{+10}_{-13}$ &$-9^{+9+4}_{-11-2}$
\\

$\bar B^0\to \bar K_0^{*-}\rho^+$ &$1^{+1}_{-2}$ &$-2^{+1+1}_{-1-0}$
&$1^{+0}_{-0}$ &$1^{+0+0}_{-0-0}$
\\

$B^-\to \bar K_0^{*-}\omega$ &$3^{+6}_{-7}$ &$-4^{+2+5}_{-3-3}$
&$-1^{+4}_{-5}$ &$-4^{+3+5}_{-4-6}$
\\

$\bar B^0\to \bar K_0^{*0}\omega$ &$16^{+26}_{-39}$
&$17^{+28+7}_{-40-5}$ &$-19^{+14}_{-15}$ &$-4^{+15+19}_{-19-13}$
\\
\hline \hline
\end{tabular}
\end{center}
\end{table}
\section{Summary} \label{sec:5}
Motivated by recent measurements of decays $B \to K_0^* \rho$ and $K_0^*
\phi$, we studied the branching fractions of these decays both in SM
and in the family nonuniversal $Z^\prime$ model within the QCDF
framework. Because it is not clear whether $K_0^*$ is the lying
state or the first orbital excited state, we calculate them under
two different scenarios. For these decay modes with scalar meson,
the weak annihilations play more important roles than that in $B\to
PP$ and  $PV$ decays, so that they will take large uncertainties. From this
point, an effective way that could  calculate the annihilations reliably is
needed. Comparing with the experimental results, we found different
channels favor different scenarios. Moreover, in order to account
for the large isospin asymmetries in the data, large weak
annihilations are also required. Adding the contribution of the
family nonuniversal $Z^\prime$ boson, we note that both the
branching fractions and their ratios are changed remarkably.
However, we cannot identify the character of the scalar meson
$K_0^*$, either. Furthermore, for most channels, the
$Z^\prime$ contribution will be buried by large uncertainties,
except for decays $B^- \to \bar K_0^{*-} \rho^0 (\omega)$.

In this work, we also calculated the  $CP$ asymmetries of these decays and found
the $CP$ asymmetries of $B\to K_0^{*}\phi$ are almost zero. In
different scenarios, the $CP$ asymmetries of $B^-\to \bar K_0^{*-}
\rho^0 (\omega)$ have different signs, thus they can be used to
classify the scalar  $K_0^{*}$. If its character is identified, we accordingly
could used these results to probe the new gauge boson $Z^\prime$,
because it changes the signs of $CP$ asymmetries. All above results could be
tested in the running LHCb or the Super-b factories in future.

\section*{Acknowledgement}
The work of Y.Li is supported by the National Science Foundation
(Nos.11175151) and the Natural Science Foundation of Shandong
Province (ZR2010AM036).


\end{document}